\documentstyle[twoside,fleqn,espcrc2,epsfig,psfig]{article}
\title{NEUTRINO MASSES AND MIXING}
\author{M. C. Gonzalez-Garcia \\
Inst. de F\'{\i}sica Corpuscular 
- C.S.I.C. - Dept. de F\'{\i}sica Te\`orica, Univ. de
Val\`encia\\
46100 Burjassot, Val\`encia, Spain}
\begin{document}
\begin{abstract}
I review the status of neutrino masses and mixings
in the light of the solar and atmospheric neutrino data.
The result from the LSND experiment and the possible role of neutrinos 
as hot dark matter are also included. I also discuss the
simplest schemes proposed to 
reconcile these data which include a light sterile neutrino 
in addition to the three standard ones. Implications for future
experiments are commented.
\end{abstract}
\maketitle
\section{Introduction}
Neutrinos are the only massless fermions predicted by the Standard Model 
(SM)\footnote{To Appear in Proceedings of Fifth International Workshop
on Tau Lepton Physics, September 1998, Santander, Spain}. This seems to be a reasonable assumption as none of the 
experiments designed to measure
the neutrino mass in laboratory experiments have found any positive
evidence for a non-zero neutrino mass. At present the existing limits
from laboratory searches are~\cite{pdg}:
\begin{eqnarray}
m_{\nu_e} & < & 15 \;\mbox{eV} \nonumber \\
m_{\nu_\mu} & < & 170 \;\mbox{KeV} \nonumber \\
m_{\nu_\tau} & < &18.2 \;\mbox{MeV} \nonumber
\end{eqnarray}
The square of the electron neutrino mass is measured in tritium beta decay
experiments by fitting the end point distribution. In several of these
experiments there has been found a negative mass squared which is
concluded to be due to unknown effects which cause the accumulation of 
events near the endpoint. This makes the limit above still far from
certain. The muon neutrino mass limit is derived from the measurement
of the muon neutrino momenta on the decay $\pi^+\rightarrow  \mu^+ \nu_\mu$,
while the tau neutrino mass limit given above is based on kinematics
of $\tau$ decays. For a detail discussion on the $\tau$ neutrino mass
limit see \cite{mutautalks}.

However, the confidence on the masslessness of the neutrino is now
under question due to the important results of underground experiments,
starting by the geochemical experiments of Davis and collaborators till 
the more recent Gallex, Sage, Kamiokande and SuperKamiokande
experiments \cite{solarexp,atmexp,superkatm98}. 
Altogether they provide solid evidence for the existence of anomalies
in the solar and the atmospheric neutrino fluxes.
Particularly relevant has been the recent confirmation by the
SuperKamiokande collaboration \cite{superkatm98} of the atmospheric
neutrino zenith-angle-dependent deficit which strongly indicates towards
the existence of $\nu_\mu$ conversion. Together with these results 
there is also the indication for
neutrino oscillations in the $\bar\nu_\mu \rightarrow \bar\nu_e$ channel 
by the LSND experiment~\cite{lsnd}. If one tries to
include all these requirements in a single framework,
we finds three mass scales
involved in neutrino oscillations. The simplest way to reconcile these
requirements invokes the existence of a light sterile neutrino 
i.e. one whose interaction with
standard model particles is much weaker
than the SM weak interaction so it does not affect the invisible Z decay 
width, precisely measured at LEP  
\cite{four}.
To this we may add the possible role of neutrinos in the dark matter
problem and structure formation \cite{dark,cobe,iras}. 

\section{Indications for Neutrino Mass}
\subsection{Solar Neutrinos}
\label{solar}
At the moment, evidence for a solar neutrino deficit comes from four
experiments \cite{solarexp}, Homestake, Kamiokande,
Gallex and Sage experiments.
The most recent data on the rates can be summarized as: 
\begin{eqnarray}
\mbox{Clorine} & & 2.56 \pm 0.23\;\; \mbox{SNU} \nonumber \\ 
\mbox{Gallex and Sage} &  & 72.2 \pm
5.6 \;\;\mbox{SNU} \nonumber \\
\mbox{Superkamiokande} & & 
(2.44 \pm 0.10) \times 10^6\;\;\mbox{cm$^{-2}$s$^{-1}$} \nonumber
\end{eqnarray}
The different experiments are sensitive to different parts of the 
energy spectrum of solar neutrinos and
putting all these results together seems to 
indicate that the solution to the problem is not astrophysical but must
concern the neutrino properties. Moreover, non-standard astrophysical 
solutions are strongly constrained
by helioseismology studies \cite{Bahcall98,helio97}. Within the
standard solar model approach, the theoretical predictions clearly lie
far from the best-fit solution what leads
us to conclude that new particle physics is the only
way to account for the data.
\begin{figure}
\centerline{
\protect\hbox{\psfig{file=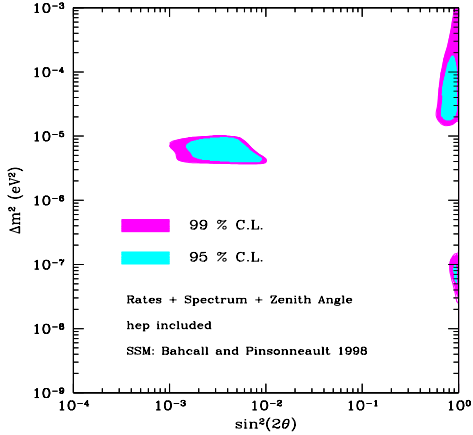,width=8cm,height=7cm}}}
\caption{Presently allowed MSW solar neutrino parameters for 2-flavour
active neutrino conversions with an enhanced $hep$  flux, from
Ref.~\protect\cite{bkhep}}
\label{msw}
\end{figure}

The standard explanation for this deficit
would be the oscillation of $\nu_e$ to another neutrino species either active
or sterile. Different analyses have been performed to find the allowed mass
differences and mixing angles in the two-flavour approximation
\cite{sol2fam,bks98,bkhep}.
The last result from Refs.~\cite{bks98,bkhep} indicate 
that for oscillations into active neutrinos there are 
three possible solutions for the parameters:\\
\begin{itemize}
\item [$\bullet$] vacuum (also called ``just so'') oscillations with 
$\Delta m^2_{ei}=(0.5$--$8)\times 10^{-10}$  eV$^2$ and $\sin^2(2\theta)=0.5$--$1$ 
\item[$\bullet$] non-adiabatic-matter-enhanced oscillations via the MSW mechanism \cite{msw} 
with $ \Delta m^2_{ei}=(0.4$--$1)\times 10^{-5}$ eV$^2$ and 
$\sin^2(2\theta)=(1$--$10)\times 10^{-3} $, and 
\item[$\bullet$] large mixing via the MSW mechanism with 
$\Delta m^2_{ei}=(0.3$--$3)\times 10^{-4}$  eV$^2$ and 
$\sin^2(2\theta)=0.6$--$1$.
\end{itemize}
\begin{figure}
\centerline{\protect\hbox{\psfig{file=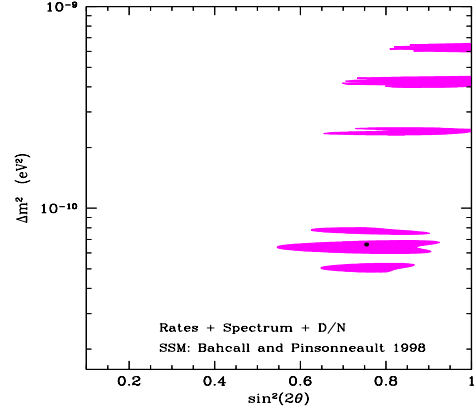,width=8cm,height=6.6cm}}}
\caption{Presently allowed vacuum oscillation
parameters, from Ref.~\protect\cite{bks98}}
\label{vac98}
\end{figure}
In Fig.~\ref{msw} I show the allowed two-flavour regions obtained in an
updated MSW global fit analysis of the solar neutrino data for the
case of active neutrino conversions. 
The analysis uses the model from 
\cite{BP98} but with an arbitrary $hep$ (from the reaction 
${\rm ^3He} ~+~ p 
~\rightarrow ~{\rm ^4He} ~+~e^+ ~+~\nu_e $)
neutrino flux ~\cite{bks98}.

Fig.~\ref{vac98} shows the regions of just-so oscillation parameters
obtained in a recent global fit of the data. It has been 
pointed out that the expected seasonal effects in this scenario
(due to the variation of the Earth-Sun distance) could
be used to further constrain the parameters \cite{lisi}, 
and also to help discriminating it from the MSW transition.

For oscillations into an sterile neutrino there are differences partly due to
the fact that now the survival probability depends both on the electron and
neutron density in the Sun but mainly due to the lack of neutral current
contribution to the Kamiokande experiment. This last effect requires a larger
$\nu_e$ survival probability. As a result the vacuum oscillation solution is
very marginal and the large mixing MSW solution is ruled out. The small mixing
solution is still valid \cite{bks98,sterile}. 

The large mixing solution for oscillations into sterile neutrinos is also in
conflict with the constraints from  big bang nucleosynthesis (BBN) 
\cite{BBN}. The presence of
additional weakly interacting light particles, such as a light
sterile neutrino, is constrained by BBN since the $\nu_s$ would enter
into equilibrium with the active neutrinos in the early Universe
via neutrino oscillations. 
However the derivation of the BBN bounds may be  subject to large
systematical uncertainties .
For example, it has been
argued in~\cite{sarkar} that present observations of primordial Helium
and deuterium abundances can allow up to $N_\nu = 4.5$ neutrino
species if the baryon to photon ratio is small. The presence of a relic 
lepton number asymmetry in the early universe may also relax this
constraint \cite{volkas}.

\subsection{Atmospheric Neutrinos}
Atmospheric showers are initiated when primary cosmic rays hit the
Earth's atmosphere. Secondary mesons produced in this collision,
mostly pions and kaons, decay and give rise to electron and muon
neutrino and anti-neutrinos fluxes \cite{reviewatm}.  There has been a
long-standing anomaly between the predicted and observed $\nu_\mu$
$/\nu_e$ ratio of the atmospheric neutrino fluxes
\cite{atmexp}. Although the absolute individual $\nu_\mu$ or $\nu_e$
fluxes are only known to within $30\%$ accuracy, different authors
agree that the $\nu_\mu$ $/\nu_e$ ratio is accurate up to a $5\%$
precision. In this resides our confidence on the atmospheric neutrino
anomaly (ANA), now strengthened by the high statistics sample
collected at the Super-Kamiokande experiment \cite{superkatm98}.  
The most important feature of the atmospheric neutrino
535-day~ data~sample reported by the SK~collaboration at
Neutrino~98~\cite{superkatm98} is that it exhibits a {\sl
zenith-angle-dependent} deficit of muon neutrinos which is
inconsistent with expectations based on calculations of the
atmospheric neutrino fluxes. 
This experiment has marked a turning point in the significance of the
ANA. 
\begin{figure}
\centerline{\protect\hbox{\psfig{file=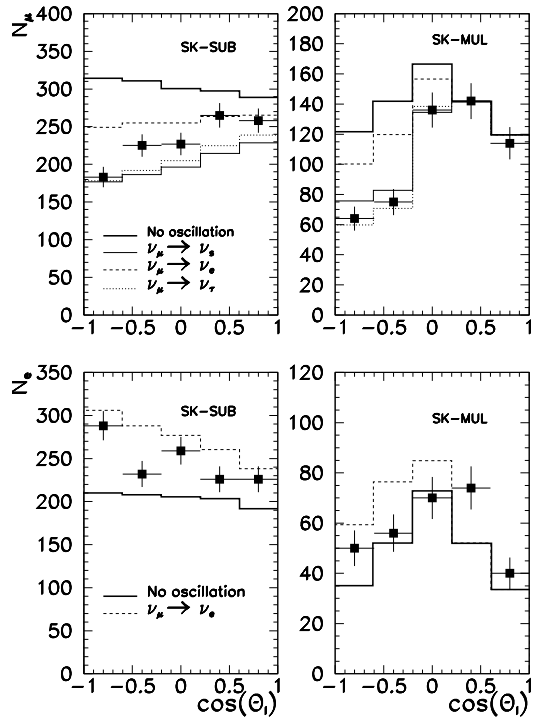,height=0.5\textheight}}}
\vglue -0.5cm
\caption{Theoretically expected zenith angle distributions for SK
electron and muon-like sub-GeV and multi-GeV events in the SM
(no-oscillation) and for the best-fit points of the various
oscillation channels, from Ref.~\protect\cite{atmours}. The
data points correspond to the 535 days of data taken from 
Superkamiokande \protect\cite{superkatm98}. }
\label{ang_mu}  
\end{figure}

In Fig.~\ref{ang_mu} I show the measured zenith angle distribution of
electron-like and muon-like sub-GeV and multi-GeV events, as well as
the one predicted in the absence of oscillation. I also give the
expected distribution in various neutrino oscillation schemes.
The thick-solid histogram is the theoretically expected distribution
in the absence of oscillation, while the predictions for the best-fit
points of the various oscillation channels is indicated as follows:
for $\nu_\mu \to \nu_s$ (solid line), $\nu_\mu \to \nu_e$ (dashed
line) and $\nu_\mu \to \nu_\tau$ (dotted line).  The error displayed
in the experimental points is only statistical.

In the theoretical analysis it has been used the latest improved
calculations of the atmospheric neutrino fluxes as a function of
zenith angle, including the muon polarization effect and took into
account a variable neutrino production point \cite{flux}.  Clearly the
data are not reproduced by the no-oscillation hypothesis, adding
substantially to our confidence that the atmospheric neutrino anomaly
is real.

\begin{figure}
\centerline{\protect\hbox{\psfig{file=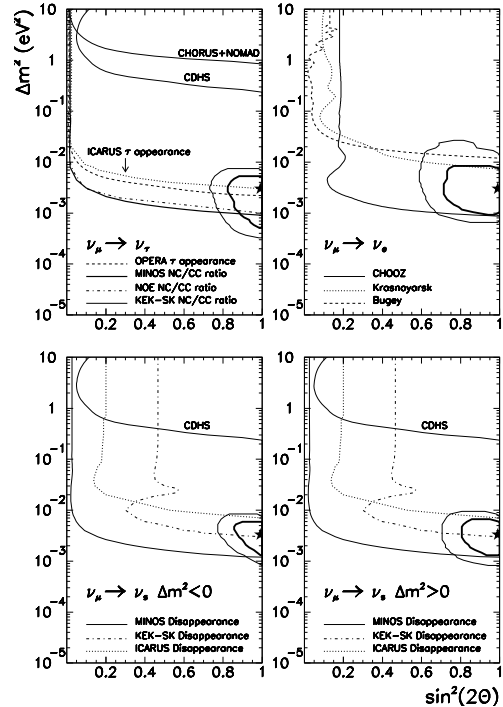,height=0.5\textheight}}}
\vglue -0.5cm
\caption{Allowed atmospheric oscillation parameters for all
experiments including the SK data reported at Neutrino~98, combined at
90 (thick solid line) and 99 \% CL (thin solid line) for all possible
oscillation channels, from Ref.~\protect\cite{atmours}.  
The sensitivity of the present
accelerator and reactor experiments as well as the expectations of
upcoming long-baseline experiments is also displayed.}
\label{mutausk4} 
\end{figure}

The most likely solution of the ANA involves neutrino
oscillations \cite{osciat}.  
In principle we can invoke various neutrino oscillation
channels, involving the conversion of $\nu_\mu$ into either $\nu_e$ or 
$\nu_\tau$
(active-active transitions) or the oscillation of $\nu_\mu$ into a sterile
neutrino $\nu_s$ (active-sterile transitions) \cite{atmours,atmothers}. 
In Fig.~\ref{mutausk4} I show the allowed neutrino oscillation parameters
obtained in a recent global fit of the sub-GeV and multi-GeV
(vertex-contained) atmospheric neutrino data~\cite{atmours}
including the recent data reported at Neutrino~98, as well as all
other experiments combined at 90 (thick solid line) and 99 \% CL (thin
solid line) for each oscillation channel considered.
The two lower panels in Fig.~\ref{mutausk4} differ in the sign of the $\Delta
m^2$ which was assumed in the analysis of the matter effects in the
Earth for the $\nu_\mu \to \nu_s$ oscillations. It was  found that $\nu_\mu
\to \nu_\tau$ oscillations give a slightly better fit than $\nu_\mu
\to \nu_s$ oscillations.  At present the atmospheric neutrino data
cannot distinguish between the $\nu_\mu\rightarrow\nu_\tau$ and $\nu_\mu \rightarrow \nu_s$ channels. Notice that in all channels where matter effects play a role 
the range of acceptable $\Delta m^2$ is
shifted towards larger values, when compared with the $\nu_\mu \to
\nu_\tau$ case. This follows from looking at the relation between 
mixing {\sl in vacuo} and in matter. In fact, away from the
resonance region, independently of the sign of the matter potential,
there is a suppression of the mixing inside the Earth. As a result,
there is a lower cut in the allowed $\Delta m^2$ value, and it lies
higher than what is obtained in the data fit for the $\nu_\mu \to
\nu_\tau$ channel.  

I also display in Fig.~\ref{mutausk4} the sensitivity of present
accelerator and reactor experiments, as well as that expected at
future long-baseline (LBL) experiments in each channel.  The first
point to note is that the Chooz reactor \cite{chooz} data already
excludes the region indicated for the $\nu_{\mu} \rightarrow \nu_e$ channel
when all experiments are combined at 90\% CL.
From the upper-left panel in Fig.~\ref{mutausk4} one sees that the regions
of $\nu_\mu \to \nu_\tau$ oscillation parameters obtained from the
atmospheric neutrino data analysis cannot be fully tested by the LBL
experiments, as presently designed.  One might expect that, due to the
upward shift of the $\Delta m^2$ indicated by the fit for the sterile
case (due to the effects of matter in the Earth) it would be possible
to completely cover the corresponding region of oscillation
parameters. Although this is the case for the MINOS disappearance
test, in general most of the LBL experiments can not completely probe
the region of oscillation parameters allowed by the $\nu_\mu \to
\nu_s$ atmospheric neutrino analysis.  This is so irrespective of the
sign of $\Delta m^2$ assumed.  For a discussion of the various
potential tests that can be performed at the future LBL experiments in
order to unravel the presence of oscillations into sterile channels
see Ref.~\cite{atmours}.
\subsection{LSND}
\begin{figure}
\centerline{\protect\hbox{\psfig{file=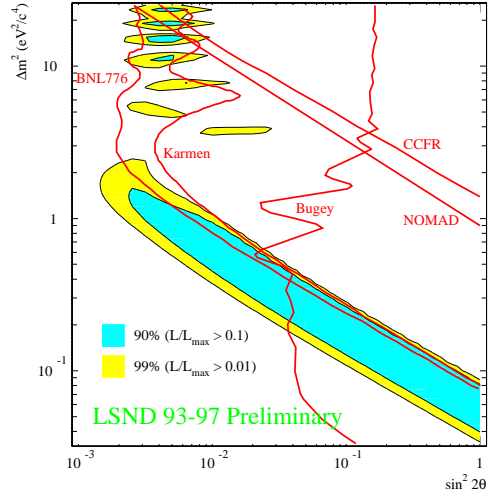,height=0.35\textheight}}}
\vglue -0.5cm
\caption{Allowed LSND oscillation parameters compared with the 90 \%
exclusion regions from other experiments.}
\label{lsnd} 
\end{figure}

Los Alamos Meson Physics Facility  (LSND) has searched  
for $\bar\nu_{\mu}\to \bar\nu_{e}$ oscillations with 
$\bar\nu_\mu$ from $\mu^+$ decay at rest \cite{lsnd}. 
The $\bar\nu_e$'s are detected in the quasi elastic process 
$\bar\nu_e\,p \to e^{+}\,n$ in correlation with a monochromatic 
photon of $2.2$  MeV arising
from the neutron capture reaction $np \to d \gamma$. In Ref.~\cite{lsnd}
they report a total of 22 events with $e^+$ energy
between 36 and $60$ MeV while  $4.6 \pm 0.6$ background
events are expected. They fit the full $e^+$ event sample in the energy
range $20 <E_e<60$ MeV by a $\chi^2$ method and the result yields
$64.3^{+18.5}_{-16.7}$ beam-related events. Subtracting the estimated
neutrino background with a correlated gamma of $12.5\pm 2.9$ events
results into an excess of 
$51.8^{+18.7}_{-16.9} \pm 8.0$ events. The interpretation of this
anomaly in terms of
$\bar \nu_\mu \to \bar \nu_e$ oscillations leads to an 
oscillation probability of ($0.31^{+0.11}_{-0.10}
\pm 0.05$)\%. 
Using a likelihood method they obtain a consistent result
of ($0.27^{+0.12}_{-0.12} \pm 0.04$)\%. 
In the two-family formalism this result  
leads to the oscillation parameters shown in
Fig.~\ref{lsnd}. The shaded regions are the 
90~\% and 99~\%  likelihood regions from LSND. Also shown are the limits from
BNL776, KARMEN1, Bugey, CCFR, and NOMAD.
\vskip .2cm
\subsection{Dark Matter}
\vskip .1cm
There is increasing evidence that more than $90\%$ of the mass in the Universe
is dark and non-baryonic. Neutrinos, if massive,  constitute a source for dark
matter. Stable neutrinos can fill the Universe of hot dark matter if their
masses add up to a maximum of about 30 eV.  
However, scenarios with only hot dark
matter run into trouble in the explanation of the formation of structures 
on small scales of the Universe.  
The research on the nature of the cosmological dark matter and the
origin of galaxies and large scale structure in the Universe within
the standard theoretical framework of gravitational collapse of
fluctuations as the origin of structure in the expanding universe has
undergone tremendous progress recently. Indeed the observations of
cosmic background temperature anisotropies on large scales performed
by the COBE satellite \cite{cobe} combined with cluster-cluster
correlation data e.g. from IRAS~\cite{iras} cannot be reconciled with
the simplest cold dark matter (CDM) model.

Currently, the best scenario for a zero cosmological constant 
to explain the data considers a mixture of cold plus hot dark matter 
\cite{dark}. This  translates into an upper limit on neutrino
masses:
\begin{equation}
\sum_i m_{\nu_i}< \mbox{few eV}\; .
\end{equation}
This mass scale is similar to
that indicated by the hints reported by the LSND experiment
\cite{lsnd}.

Very recent data on Type-I supernovae at high redshifts \cite{cosmoconst}
has provided evidence at more than 99 \% CL for an accelerating expansion
of the universe. They measure the light curve of the supernovae 
which gives the absolute luminosity. In this way they are able to determine 
the distance as a function of the redshift, and from there to measure the
deceleration parameter $q_0=\Omega_M/2-\Omega_\Lambda$. They find
$q_0<0$. $q0$ gives a measurement of the different contributions to the 
energy density in the universe coming from matter and from the presence of 
a cosmological constant. In a flat universe both contributions must verify 
$\Omega_M+\Omega_\Lambda$=1. In other words, the data from supernovae 
searches indicate a non-zero cosmological constant, or equivalently, 
$\Omega_M < 1$. Actually the results indicate, for a flat universe, 
$\Omega_M < 0.5$ at 99 \% CL.

Should these results be confirmed the amount of dark matter of the universe
would be considerably reduced and in consequence the corresponding limit
on the stable neutrino mass will become tighter. 
Future sky maps of the cosmic microwave background radiation (CMBR)
with high precision at the upcoming MAP and PLANCK missions should
bring more light into the nature of the dark matter and the possible
role of neutrinos.

\section{Reconciling the neutrino puzzles}

Naive two-family counting shows that it is very difficult to  fit
all experimental information even in the three neutrino scenario, 
even without invoking the LSND data. One has to choose
between throwing away part of the data and considering a larger scheme.

The solar neutrino deficit could be due to $\nu_e\rightarrow \nu_\mu$ 
oscillations and the atmospheric neutrino deficit to $\nu_\mu\rightarrow
\nu_\tau$ oscillations with the appropriate mass differences, for example with
a mass hierarchy  $m_{\nu_\tau}\gg m_{\nu_\mu}, m_{\nu_e}$. However, fitting
this together with the present laboratory limits leaves no room for hot dark
\cite{fogli}. The only possible way out is to require that all three neutrinos
are almost degenerate.  This requires a certain degree of fine-tuning in order
to explain the neutrinoless double beta decay data.
Notice that this scenario is also inconsistent with the  oscillation
parameters observed by LSND.

One could have $\nu_\mu\rightarrow \nu_\tau$ oscillations for the atmospheric 
neutrino deficit with  almost degenerate $\nu_\mu$ and $\nu_\tau$ with masses
$m_{\nu_\mu}=m_{\nu_\tau}\approx$ few eV and $m_{\nu_e}\approx 0$,  
but leaving out the explanation for the solar neutrino
deficit.  Or $m_{\nu_\mu}=m_{\nu_\tau}\approx 0$  and 
$m_{\nu_e}\approx $ few eV
to explain the atmospheric data but leaving unexplained  both solar neutrino
deficit and dark matter. 
 
Also, it is possible to explain the solar neutrino deficit with
$\nu_e\rightarrow \nu_{\tau(\mu)}$ with almost degenerate $\nu_e$ and
$\nu_{\tau(\mu)}$ with masses   $m_{\nu_e}=m_{\nu_{\tau(\mu)}}\approx$ few eV  
and $m_{\nu_{\mu (\tau)}}\approx 0$, but leaving the atmospheric neutrino
deficit  unexplained. Also
$m_{\nu_e}=m_{\nu_{\tau}}\approx 0$ and 
$m_{\nu_{\mu}} \approx$ eV would explain the LSND data if 
confirmed but leaves  both atmospheric and dark matter without explanation.

The ``minimal" scheme to explain $all$ data without fine-tuning seems to be a
four-neutrino framework $(\nu_e,\nu_\mu,\nu_\tau,\nu_s)$ where $\nu_s$ is a
sterile neutrino. 

\subsection{Four-Neutrino Models}

The simplest way to open the possibility of incorporating the LSND
scale to the solar and atmospheric neutrino scales is to invoke a sterile 
neutrino, i.e. one whose interaction with
standard model particles is much weaker
than the SM weak interaction so it does not affect the invisible Z decay 
width, precisely measured at LEP. The sterile neutrino must also be light
enough in order to participate in the oscillations involving the three
active neutrinos \cite{four}. 

\begin{figure}
\centerline{\protect\hbox{\epsfig{file=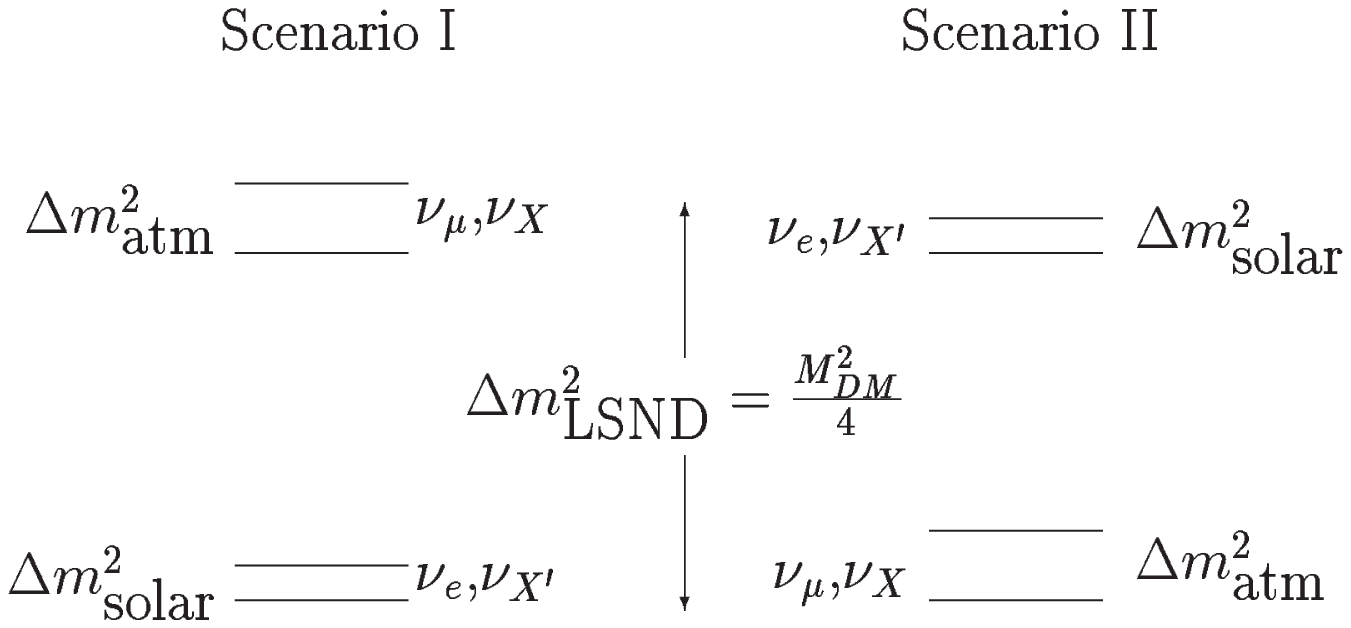,width=0.5\textwidth,height=0.25\textheight}}}
\caption{Allowed scenarios for four neutrino oscillations.}
\label{masses} 
\end{figure}

After imposing the present constrains from the negative searches
at accelerator \cite{accelerator} and reactor \cite{chooz,reactor} 
neutrino oscillation experiments
one is left with two possible mass patterns as described in 
Fig.~\ref{masses} which I will call scenario I and II.
In scenario I there are two lighter neutrinos at the solar neutrino
mass scale and two maximally mixed almost degenerate eV-mass
neutrinos split by the atmospheric neutrino scale.
In scenario II the two lighter neutrinos are maximally mixed and
split by the atmospheric neutrino scale while the two
heavier neutrinos are almost degenerate separated by the solar
neutrino mass difference. In both scenarios solar neutrino data
together with reactor neutrino constrains, imply that the electron 
neutrino must be maximally projected over one of the states 
belonging to the pair split by the solar neutrino scale: the 
lighter (heavier) pair for scenario I (II). On the other hand,
atmospheric neutrino data together with the bounds from accelerator neutrino 
oscillation experiments imply that the muon neutrino must be maximally 
projected over the pair split by the atmospheric neutrino mass difference:
the heavier (lighter) pair for scenario I (II).

In both scenarios there are two possible assignments for the sterile and 
tau neutrinos which I denote by .a and .b depending on whether the tau neutrino
is maximally projected over the pair responsible for the atmospheric
neutrino oscillations and the sterile neutrino is responsible for
the solar neutrino deficit ($\nu_X=\nu_\tau$ and $\nu_{X^\prime}=\nu_s$) 
or viceversa ($\nu_X=\nu_s$ and $\nu_{X^\prime}=\nu_\tau$).

These four possibilities offer different signatures at future
experiments:
\begin{itemize} 
\item In scenario II the electron neutrino must be mainly composed
of one of the heavier states with a mass characteristic of the
LSND mass difference and the dark matter 
$m_1=m_{DM}/2=\sqrt{\Delta m^2_{LSND}}\sim 0.1--\mbox{few eV}$ and
may be tested at future neutrinoless double-$\beta$ decay and tritium
$\beta$ decay experiments.
\item As mentioned before in the atmospheric neutrino section, 
Scenarios I.a and II.a give a slightly better fit to the atmospheric neutrino
anomaly. Future long baseline experiments can be sensitive to this
oscillation and the most sensitive test would be a $\tau$ appearance
experiment.
\item Scenarios I.b and II.b where the atmospheric $\nu_\mu$ deficit
is due to the oscillation into an sterile neutrino imply a higher 
value of $\Delta m^2_{atm}$ as can be seen in Fig.~\ref{mutausk4} due
to the effect of propagation through the Earth which suppresses
the lower mass region. As a consequence this scenario can be
easier to test at future long baseline experiments. However
only a disappearance-type experiment is possible and, in general
these tests can achieve lower sensitivity. 
\item For solar neutrinos the three regions discussed in subsection \ref{solar}
are valid for scenarios I.a and II.a when the solar data is
explained in terms of $\nu_e\rightarrow \nu_\tau$. For scenarios I.b and 
II.b where $\nu_e\rightarrow \nu_s$ are invoked to account for 
the solar neutrino deficit, there are differences
mainly due to the lack of neutral current
contribution to the Kamiokande experiment. This last effect requires a larger
$\nu_e$ survival probability. As a result the vacuum oscillation solution is
very marginal and the large mixing MSW solution is ruled out. 

\item The neutral-to-charged current ratio is a very 
important observable in neutrino oscillation phenomenology, which is
especially sensitive to the existence of singlet neutrinos.
This test can be carried out both at future solar and atmospheric 
neutrino experiments as well as long baseline experiments. 
At present one may study the ratios of $\pi^0$-events and the events
induced mainly by the charged currents ~\cite{vissani}. 
Superkamiokande has reported the result \cite{superkatm98}
\begin{displaymath}
\frac{(\pi^0/e)_{data}}{(\pi^0/e)_{MC}}=0.93\pm 0.07 \pm 0.19
\end{displaymath}
The expected values are 1. (0.75) for scenarios I.a and II.a  (I.b and II.b).
The result above  is consistent with both scenarios with a
slight preference for the former. 

\item In scenarios I.b and II.b the sterile neutrino is largely mixed
with one active neutrino. This gives a larger contribution to the 
effective degrees of freedom at the time of BBN. 
Should the BBN constrains
become more precise, these scenarios may be ruled out. 
\end{itemize}

\section{Conclusions}

The impressive re-confirmation of an angle-dependent
atmospheric neutrino deficit by 
Superkamiokande leaves little room for doubt that the
atmospheric neutrino anomaly is an strong evidence for
neutrino masses and mixings. 
Likewise, it has become more and more difficult to avoid neutrino
oscillations as an explanation for the solar neutrino
puzzle. Also the LSND evidence for
$\bar\nu_e$-$\bar\nu_\mu$-oscillations still remains a viable hypothesis
although more restricted by the exclusion limit of the KARMEN 
experiment. These three results can be interpreted in terms of
neutrino oscillations but with the need of three different mass
scales. Thus if the LSND result stands the test of time, this would be a 
puzzling indication for the existence of a light sterile neutrino.

The two scenarios to reconcile these observations invoke
either $\nu_e\rightarrow \nu_\tau$ oscillations to explain the solar data, 
with $\nu_\mu\rightarrow\nu_s$ 
oscillations accounting for the atmospheric deficit, or viceversa. 
They have distinct implications 
at future tritium beta decay and neutrino-less double beta decay 
experiments as well as solar, atmospheric and long baseline neutrino 
experiments. In particular the neutral-to-charged current ratio is an
important observable to discriminate among the different scenarios
as it is sensitive to the existence of singlet neutrinos.
This test can be carried out both at future solar and atmospheric 
neutrino experiments as well as long baseline experiments.
\vskip 1cm
\noindent
{\bf Acknowledgments} \\
\noindent
I am grateful to A. Pich and A. Ruiz  for the kind 
hospitality in Santander. My thanks to John Bahcall and Bill Louis, 
for providing me with  their postscript figures. 
This work was supported by
grants CICYT AEN96-1718, DGICYT PB95-1077 and  DGICYT PB97-1261, and
by the EEC under the TMR contract ERBFMRX-CT96-0090.

\end{document}